\documentclass[fleqn,usenatbib]{mnras}
\usepackage{newtxtext,newtxmath}
\usepackage[T1]{fontenc}
\usepackage{ae,aecompl}

\usepackage{graphicx,caption}
\usepackage{amsmath}	    
\usepackage[dvipsnames]{xcolor}

\graphicspath{{./plots/}}

\title[GL of GWs: Probing IMBHs in galaxy lenses]{Gravitational lensing of gravitational waves: prospects for probing intermediate-mass black holes in galaxy lenses with global minima image}

\author[A. K. Meena]{
Ashish Kumar Meena$^{1}$\thanks{E-mail: \href{mailto:ashishmeena766@gmail.com}{ashishmeena766@gmail.com}}
\\
\\
$^{1}$Physics Department, Ben-Gurion University of the Negev, P.O. Box 653, Be'er-Sheva 84105, Israel 
}

\date{Accepted XXX; Received YYY; in original form ZZZ}

\pubyear{2024}

\begin{document}
\label{firstpage}
\pagerange{\pageref{firstpage}--\pageref{lastpage}}
\maketitle

\begin{abstract}
This work studies microlensing effects in strongly lensed gravitational wave~(GW) signals corresponding to global minima images in galaxy-scale lenses. We find that stellar microlenses alone are unable to introduce noticeable wave effects in the global minima GW signals at strong lensing magnification~$(\mu)<50$ with match value between unlensed and lensed GW signals being above~${\sim}99.5\%$ in ${\sim}90\%$ of systems implying that GW signals corresponding to global minima can be treated as reference signal to determine the amount of microlensing in other strongly lensed counterparts. Since the stellar microlenses introduce negligible wave effects in global minima, they can be used to probe the intermediate-mass black hole~(IMBH) lenses in the galaxy lens. We show that the presence of an IMBH lens with mass in the range~$[50,10^3]~{\rm M_\odot}$ such that the global minima lies within five Einstein radius of it, the microlensing effects at~$f<10^2$~Hz are mainly determined by the IMBH lens for~$\mu<50$. Assuming that a typical strong lensing magnification of~3.8 and high enough signal-to-noise ratio~(in the range~${\simeq}[10, 30]$) to detect the microlensing effect in GW signals corresponding to global minima, with non-detection of IMBH-led microlensing effects in ${\simeq}15~({\simeq}150)$ lensed GW signals, we can rule out dark matter fraction~$>10\%~(>1\%)$ made of IMBH population inside galaxy lenses with mass values~$>150~{\rm M_\odot}$ with~${\sim}$90\% confidence. Although we have specifically used IMBHs as an example, the same analysis applies to any subhalo~(or compact objects) with lensing masses~(i.e., the total mass inside Einstein radius) satisfying the above criterion.
\end{abstract}


\section{Introduction}
\label{sec:intro}

Following the first detection of gravitational waves~\citep[GWs;][]{2016PhRvL.116f1102A} by Laser Interferometer Gravitational-Wave Observatory~\citep[LIGO;][]{2015CQGra..32g4001L}, one of the highly anticipated observation is a lensed GW signal. So far, the LIGO-Virgo-KAGRA~\citep[LVK;][]{2012CQGra..29l4007S, 2015CQGra..32b4001A} detector network has detected a total of~90 GW signal coming from binary black hole~(BH), binary neutron star~(NS), and BH-NS mergers in the first three runs~(O1, O2, O3a, and O3b;~\citealt{2021arXiv211103606T}). None of these events are proved to be lensed~\citep[e.g.,][]{2019ApJ...874L...2H, 2021ApJ...908...97L, 2021ApJ...923...14A, 2023arXiv230408393T}. However, with an increase in the sensitivity of current detectors in the upcoming runs and new detectors coming online (Cosmic Explorer:~\citealt{2021arXiv210909882E}; DECIGO:~\citealt{2021PTEP.2021eA105K}; Einstein Telescope:~\citealt{2020JCAP...03..050M}; LISA:~\citealt{2020GReGr..52...81B}), the detection of a lensed GW signal is expected to become a reality in the near future~\citep[e.g.,][]{2018MNRAS.476.2220L, 2022ApJ...929....9X}.

The estimated remnant BH masses of mergers detected in LIGO lie in a range from~${\sim}10~{\rm M_\odot}$ to~${\sim}200~{\rm M_\odot}$ starting to reveal the population of intermediate-mass BHs~(IMBHs) with mass~$>100~{\rm M_\odot}$~\citep[e.g,][]{2020PhRvL.125j1102A}. Detection of such IMBHs bridges the observed gap between stellar mass BHs (with masses of a few tens of~${\rm M_\odot}$) and super-massive BHs in mass range~$\sim[10^6, 10^{10}]~{\rm M_\odot}$ and allow us to understand their formation channels~\citep[e.g.,][]{2017IJMPD..2630021M, 2020ARA&A..58..257G}~\citep[e.g.,][]{2017IJMPD..2630021M, 2020ARA&A..58..257G} as well as the existence of super-massive BHs in the early Universe~\citep[e.g.,][]{2018Natur.553..473B, 2023ApJ...953L..29L, 2024NatAs...8..126B, 2024Natur.627...59M}. At present, the formation and growth of IMBHs remain uncertain, and further understanding can be gained by detecting more such GW events where the end product is an IMBH~\citep[e.g.,][]{2020ApJ...900L..13A}. Observations suggest that IMBHs are expected to reside at the centre of dwarf galaxies~\citep[][]{2015ApJ...809L..14B} or off-centred as hyper/ultra-luminous x-ray sources~\citep[e.g.,][]{2019ApJ...882..181B} or inside globular clusters~\citep[e.g.,][]{2005ApJ...634.1093G}. IMBHs at the centre of dwarfs expect to emit electromagnetic~(EM) signatures at different frequencies, whereas IMBHs in globular clusters can be detected by studying the motion of stars or in GW observations. A non-zero population of IMBHs is expected to reside in the galaxy halos~(typically known as wandering IMBHs), which had been accreted during the hierarchical mergers of satellite galaxies or ejected from globular clusters~\citep[e.g.,][]{2014ApJ...780..187R, 2014MNRAS.441..809W, 2020ARA&A..58..257G, 2021ApJ...917...17G, 2023MNRAS.525.1479D, 2024arXiv240412354U}. Such IMBHs are expected to be naked~(no associated stellar component) or in very compact star clusters~(stellar component mass is only at per cent level) and can remain undetected in EM observations. Since gravitational lensing by stellar BHs and IMBHs has the ability to introduce frequency-dependent effects in GW signals in the LIGO frequency band~\citep[e.g.,][]{2018PhRvD..98j3022C, 2018PhRvD..98h3005L, 2022ApJ...932L...4G, 2022MNRAS.509.1358U}, lensing comes as an alternative tool which can help us to probe the population of these wandering IMBHs lying within distance galaxies, or more generally, any compact object with similar mass values. However, for convenience, throughout this manuscript, we only use IMBH to refer to such objects.

Similar to electromagnetic~(EM) waves, GW signals are also subjected to gravitational lensing~\citep[e.g.,][]{1971NCimB...6..225L, 1974IJTP....9..425O}. In principle, the lensing of GWs can be studied in the same way as the lensing of EM waves. For example, the gravitational lensing of a GW signal by a galaxy or cluster of galaxies can lead to the formation of multiple copies of the same signal arriving with a time delay between them~\citep[e.g.,][]{2018IAUS..338...98S, 2021MNRAS.506.5430J}. However, differences can arise because the frequency of GWs detected in the LIGO lies in the frequency range~[$10^1$, $10^4$]~Hz, whereas the EM waves observed in different telescopes have much higher frequencies~(e.g., $\sim10^6$~Hz for radio waves and $\sim10^{18}$~Hz for x-rays). If the time delay between the two strongly lensed copies of the same GW signal is roughly equal to the period of the GW signal~(e.g., $t_d\sim1/f$), then these two lensed GW signals will interfere with each other leading to frequency-dependent features in the detected signal~\citep[e.g,][]{1981Ap&SS..78..199B, 1986ApJ...307...30D, 1998PhRvL..80.1138N, 1999PhRvD..59h3001B, 1999PThPS.133..137N, 2003ApJ...595.1039T}. For the EM waves with frequency in the visible range~(i.e., $f\sim5\times10^{14}$~Hz), the time delay should be~$\sim10^{-15}$~seconds to observe the interference~\citep[e.g,][]{1995ApJ...442...67U}. In addition to this extremely small time delay, as most of the EM sources are not phase coherent, these interference effects will be washed out. On the other hand, in the LIGO frequency range, the time delay between two lensed GW signals needs to be~$\sim[10^{-1}, 10^{-4}]$~seconds to observe the interference effects. In galaxy or cluster scale lenses, having such short time delays is very rare but microlensing due to stellar mass objects always leads to the formation of micro-images with time delay~$\sim[10^{-6},10^{-2}]$~seconds required for interference in the LIGO frequency band~\citep[e.g,][]{2020MNRAS.492.1127M}. Inside a lens galaxy, instead of a single microlens, we have a whole population of microlenses with different lens masses, implying that the interference effects in the strongly lensed GW signals may be very complex, and one cannot model the effects using an isolated point mass lens. These effects are further sensitive to the type of strongly lensed image, and the microlensing effects may vary depending on the strong lens magnification and microlens mass function~\citep[e.g.,][]{2019A&A...627A.130D, 2021MNRAS.508.4869M, 2022MNRAS.517..872M}. However, as shown in~\citet{2022MNRAS.517..872M}, at strong lensing magnification~(or macro-magnification)~$<10$, wave effects due to typical stellar microlenses are not significant and are not expected to affect the parameter estimation.

Keeping the non-significant wave effects by stellar population in mind, we can ask: can we use strongly lensed GW signals to probe the wandering IMBH population (or other similar mass compact objects) in the strong lensing galaxies, which can be extremely hard to detect otherwise? One can answer this question by quantifying the effects of an IMBH on a strongly lensed GW signal and stellar microlensing effects on the IMBH-led lensing, as done in the current work. Following~\citet{2022MNRAS.517..872M}, in our current work, we focus on the microlensing of GW signals corresponding to global minima in galaxy-scale lenses by stellar microlensing along with an IMBH lens. In our current work, we only focus on global minima, considering the fact that at global minima, stellar microlens density will be least compared to other images implying minimal effects from stellar microlenses. Since wave effects can show very complex behaviour, by minimizing the contribution from stellar microlens by choosing the global minima, we can make sure that wave effects from IMBHs will stand out to the fullest extent. In addition, global minima is also not expected to get de-magnified, similar to saddle-points. We use \textit{match} between unlensed and lensed GW signals to quantify the effects of microlensing and to compare microlensing due to IMBH only and IMBH with stellar microlenses.

The current work is organised as follows. In Section~\ref{sec:basic}, we briefly re-visit the relevant basics of lensing. In Section~\ref{sec:stel_ml}, we study the effect of stellar microlenses on lensed GW signals corresponding to global minima. In Section~\ref{sec:stel_pbh_ml} and~\ref{sec:high_mag}, we study the effect of IMBH lens along with stellar microlenses on strongly lensed GW signal with macro-magnification of~$<20$ and in the range~$[20,50]$, respectively. In Section~\ref{sec:const_bh_pop}, we briefly discuss the constraints one can impose on the compact dark matter fraction with detecting strongly lensed global minima GW signals. We conclude this work in Section~\ref{sec:conclusions}. In this work, we use $H_0=70\:{\rm km\:s^{-1}\:Mpc^{-1}}$, $\Omega_{m}=0.3$, and $\Omega_{\Lambda}=0.7$.

\begin{figure*}
    \centering
    \includegraphics[height=5.5cm,width=11cm]{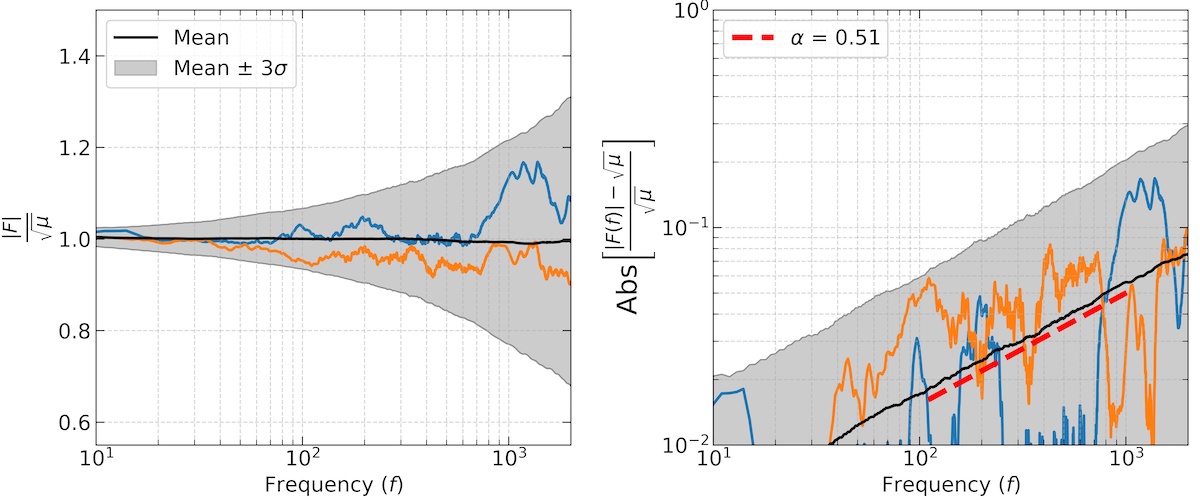}
    \includegraphics[height=5.5cm,width=6.5cm]{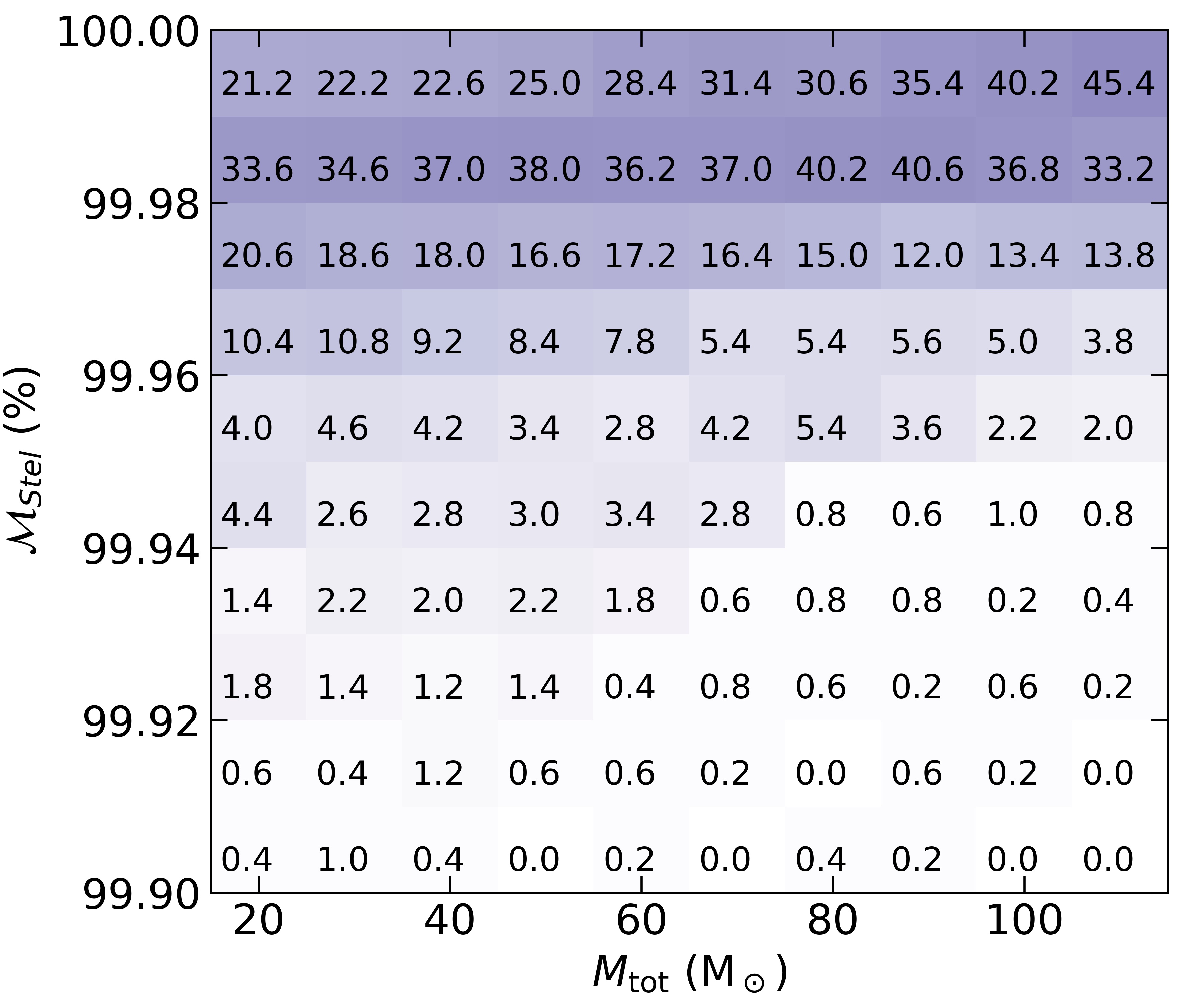}
    \caption{Effect of stellar microlenses on lensed GW signal corresponding to global minima in galaxy scale lenses. \textit{Left panel:} absolute value of normalised amplification factor~($|F|/\sqrt{\mu}$) as a function of frequency. The black solid line represents the average value from 500 realisations. The shaded region around it shows the~$3\sigma$ range around the mean value. The blue and yellow curves show the normalised amplification factor for two typical individual realisations. \textit{Middle panel:} deviation in amplification factor due to stellar microlenses from macro-magnification, i.e. relative difference. The black solid line shows the average value. The grey-shaded region covers the~$3\sigma$ range around the mean value. \textit{Right panel:} mismatch between unlensed and microlensed GW signal assuming equal mass binaries. The x-axis represents the total binary mass, with both components having equal mass. For each binary mass, (rounded) values in black colour represent the percentage of systems with match values on the y-axis.}
    \label{fig:Ff_stel}
\end{figure*}

\section{Basics theory}
\label{sec:basic}

The GW detectors measure the amplitude of the signal, unlike in the case of EM waves, where the telescopes measure the flux. Hence, in the gravitational lensing of GW signals, we estimate the amplification factor~($F$) for a lensed GW signal instead of the magnification factor~($\mu$) as we do for the lensing of EM sources. Lensing of GW sources leading to the formation of multiple copies of the same GW signal can be studied either using geometric optics or wave optics approximation. Geometric optics approximation is valid when~$f t_d \gg 1$, where~$f$ is the frequency of GW signal and~$t_d$ is the observed time-delay between the lensed GW~signals. On the other hand, when~$f t_d\sim 1$, interference between different GW signals takes place, and wave effect becomes important. In the wave optics regime, the amplification factor,~$F(f)$, is given by~\citep[e.g.,][]{1992grle.book.....S, 1999PThPS.133..137N, 2003ApJ...595.1039T},
\begin{equation}
    F(f,\mathbf{y}) = \frac{\xi_0^2 (1+z_L)}{c} \frac{D_S}{D_L D_{LS}} \frac{f}{i} 
    \int d^2 \mathbf{x} \exp\left[2\pi i f t_d(\mathbf{x},\mathbf{y})\right],
    \label{eq:Ff}
\end{equation}
where $z_L$ is the lens redshift. $D_L$, $D_{LS}$, and $D_S$ are angular diameter distances from observer to lens, from lens to source, and observer to source, respectively. $\xi_0$ is an arbitrary length scale, and~$\mathbf{y}$ and~$\mathbf{x}$ are two-dimensional vectors in the source plane and image plane (in units of~$\xi_0$), respectively, and~$t_d$ represents the arrival time surface given as,
\begin{equation}
    t_d\left(\mathbf{x},\mathbf{y}\right) = \frac{\xi_0^2 (1+z_L)}{c} \frac{D_S}{D_L D_{LS}}
    \left[\frac{|\mathbf{x}-\mathbf{y}|^2}{2} - \psi(\mathbf{x}) + \phi_m(\mathbf{y})\right],
    \label{eq:td_surface}
\end{equation}
where~$\psi(\mathbf{x})$ is the lens potential and $\phi_m(\mathbf{y})$ is a constant independent of lens properties and can be chosen in a way to simplify the calculation. In geometric optics regime~($f t_d\gg1$), the integral in Equation~\eqref{eq:Ff} becomes highly oscillatory, and only the stationary points of the integrand contribute to the~$F(f)$ values. The amplification factor can be written in the form,
\begin{equation}
    F(f,\mathbf{y}) = \sum_j\sqrt{|\mu_j|} \: \exp\left(2 \pi i f t_{d,j} - i \pi n_j\right),
    \label{eq:Ff_geo}
\end{equation}
where $\mu_j$ and $t_{d,j}$ are the macro-magnification factor and time delay of $j$-th image, respectively. $n_j$ is the Morse index with values 0, 1/2, and 1 for minima, saddle-point, and maxima, respectively. Here, we notice that for a lensed GW signal, the amplification factor~($F$) is equal to the square root of the corresponding absolute magnification factor~($|\mu|$) apart from a constant phase factor which depends on the type of the lensed image. Hence, such a constant phase factor can allow us to segregate the lensed GW signal from the unlensed ones~\citep[e.g.,][]{2020arXiv200712709D, 2021ApJ...923L...1J, 2021arXiv210409339L, 2022arXiv220206334V}.

Strong lensing of a GW source leads to the formation of multiple copies of the same GW signal arriving at different times and amplified by different factors. For galaxy scale lenses,~$f t_d \gg 1$, allowing us to use geometric optics to study these lensed GW signals. However, the presence of stellar mass microlenses inside the galaxy has the ability to further affect each of these strongly lensed GW signals and introduce wave effects. To study the effect of microlensing on these individual strongly lensed GW signals, the effect of the galaxy lens as a whole can be considered as external effects and can be described by a constant convergence~($\kappa'$) and a constant shear~($\gamma' \equiv \gamma_1' + i\gamma_2'$) determining the lensing by the overall galaxy. In such a scenario, the contribution of the overall galaxy in the lensing potential can be given as~\citep[e.g,][]{2020MNRAS.492.1127M},
\begin{equation}
    \psi_{\rm ext}(x_1,x_2) = \frac{\kappa'}{2} (x_1^2 + x_2^2)
    + \frac{\gamma_1'}{2} (x_1^2 - x_2^2) + \gamma_2' x_1 x_2.
\end{equation}
The resultant potential to study the microlensing of strongly lensed GW signal can be given as,
\begin{equation}
    \begin{split}
    \psi(x_1,x_2) = \sum_i m_i \ln\left[\sqrt{\left(x_1-x_{1i}\right)^2 + 
    \left(x_2-x_{2i}\right)^2}\right] \\
    + \frac{\kappa'}{2} (x_1^2 + x_2^2) + \frac{\gamma_1'}{2} (x_1^2 - x_2^2) 
    + \gamma_2' x_1 x_2,
    \end{split}
    \label{eq:slml}
\end{equation}
where $m_i$ and $\left(x_{1i},x_{2i}\right)$ represent the mass and position of $i$-th microlens, respectively. In our current work, we use the method described in~\citet{1995ApJ...442...67U} and~\citet{2021MNRAS.508.4869M} to calculate the amplification factor,~$F(f)$, in the presence of microlenses and refer readers to these works for further details. Other methods to determine the amplification factor are discussed in~\citet{2019A&A...627A.130D} and~\citet{2023SCPMA..6639511S}.

To quantify the lensing effects introduced by strong- and micro-lensing, we use the match between lensed and corresponding unlensed GW signal. Assuming that $h_{\rm L}$  and $h_{\rm U}$ represent the lensed and unlensed GW signals, respectively, the match~$(\mathcal{M}$; \citealt{2016CQGra..33u5004U}) is defined as,
\begin{equation}
    \mathcal{M} = \max\limits_{t_0,\phi_0} \frac{\langle h_{\rm L} | h_{\rm U} \rangle}{\sqrt{\langle h_{\rm L} | h_{\rm L} \rangle \langle h_{\rm U} | h_{\rm U} \rangle}},
    \label{eq:match}
\end{equation}
where~$t_0$ and~$\phi_0$ are arrival time and phase corresponding to~$h_{\rm U}$. The inner prodct~$\langle .|. \rangle$ is noise weighted and defined as,
\begin{equation}
    \langle h_1|h_2 \rangle = 4 {\rm Re} \int_{f_{\rm low}}^{f_{\rm high}} df \frac{h_1^*(f) h_2(f)}{S_n(f)},
    \label{eq:snf}
\end{equation}
where~$S_n(f)$ is the single-sided power spectral density~(PSD) of the detector noise. Since strong lensing alone can only introduce a constant (de-)amplification or phase factor in the lensed GW signal, Equation~\eqref{eq:match} implies that the mismatch~($1-\mathcal{M}$) between unlensed and strongly lensed GW signal will be \textit{zero}\footnote{This may not hold true in presence of precession, eccentricity, or higher-order modes in the GW signal~\citep[e.g.,][]{2021PhRvD.103f4047E, 2021PhRvD.103j4055W, 2021ApJ...923L...1J}}. Hence, a non-zero mismatch will occur when we have microlensing-induced frequency-dependent effects. A high match value~($\mathcal{M}>99\%$) generally implies that microlensing-induced effects are negligible and will go unnoticed~\citep{2014PhRvD..90f2003C}. In our current work, to estimate the match values, we use \textsc{IMRPhenomPv2} waveform approximant with a lower frequency~($f_{low}$) cut value of 20~Hz and \textsc{aLIGOZeroDetHighPower} as the PSD similar to~\citet{2022MNRAS.517..872M}. For a more detailed analysis of microlensing effects on match values, we refer the reader to~\citet{2023arXiv230611479M}.

\begin{figure*}
    \centering
    \includegraphics[scale=0.4]{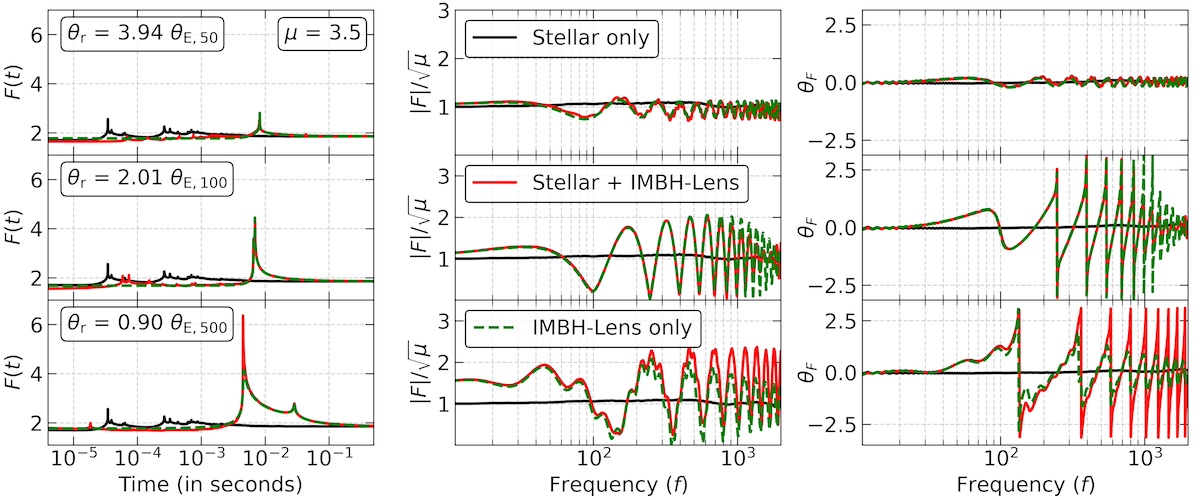}
    \caption{A typical example of the effect of IMBH on the stellar microlensing of global minima in one realisation. In \textit{top} row, \textit{left} panel shows the Fourier transform of amplification factor,~$F(t)$. Black, red, and dashed green curves are for stellar microlenses only, stellar microlenses with IMBH lens and IMBH lens only scenarios, respectively. The \textit{middle} and \textit{right} panels represent the absolute and phase of amplification factor,~$F(f)$, respectively. In \textit{top}, \textit{middle}, and \textit{bottom} rows, mass of IMBH lens is~$50~{\rm M_\odot}$, $100~{\rm M_\odot}$, and~$500~{\rm M_\odot}$, respectively. The distance of the IMBH lens from the macro-image position~($\theta_r$) is shown in the right panels.}
    \label{fig:Ff_stelBH}
\end{figure*}

\begin{figure*}
    \centering
    \includegraphics[scale=0.55]{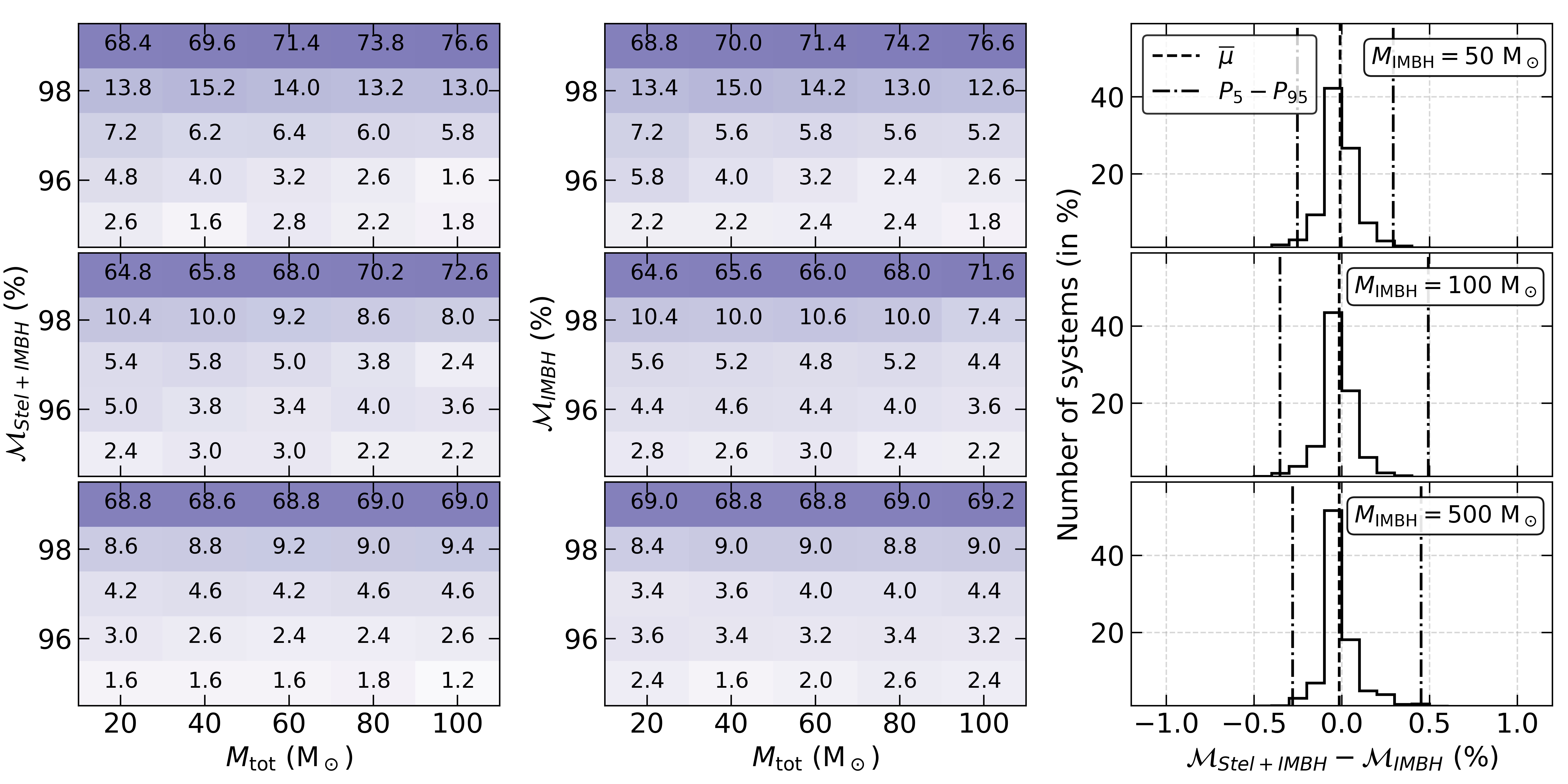}
    \caption{Effect of stellar microlenses with IMBH on match~($\mathcal{M}$) values. The \textit{top} row is for~$M_{\rm IMBH}=50~{\rm M_\odot}$. \textit{Left panel:} shows the match between unlensed and microlensed~(by stellar microlenses and IMBH;~$\mathcal{M}_{\rm Stel+IMBH}$) GW signals. \textit{Middle panel:} represents the match between unlensed and microlensed~(by IMBH only;~$\mathcal{M}_{\rm  IMBH}$) GW signals. \textit{Right panel:} shows the histogram for the difference between~$\mathcal{M}$ values shown in the left and middle panels for a match greater than~95\%. The vertical dashed line shows the median difference, and dashed-dotted lines are 5~and~95~percentiles. The \textit{middle} and \textit{bottom} rows are corresponding to~$M_{\rm IMBH}=100~{\rm M_\odot}$ and~$M_{\rm IMBH}=500~{\rm M_\odot}$, respectively.}
    \label{fig:Ff_stelBH_Match}
\end{figure*}

\section{Stellar Microlenses}
\label{sec:stel_ml}

Prior to examining the combined effects of IMBH and stellar microlenses on strongly lensed GW signals, we initially focus on studying the impact of stellar microlenses alone. A similar analysis was also done in~\citet{2022MNRAS.517..872M} for lensed GW signals with macro-magnification~$(\mu)<10$, but here we do not put any upper limit on the~$\mu$ value. For simplicity, we fix the lens redshift~($z_l$) to 0.5 and source redshift~($z_s$) to 2.0, respectively. The lens velocity dispersion and ellipticity are fixed to~250~km/s and 0.3, respectively, assuming a typical early-type galaxy~(ETG) as our lens. We model the lens galaxy using \textit{singular isothermal ellipsoid}~\citep[SIE;][]{1994A&A...284..285K} lens profile. After that, we generate strongly lensed systems by varying the source position. Once the strong lens systems are generated, we estimate the stellar surface mass density at the image position using \textit{S\'ersic} profile following eqaution~8 in~\citet{2019MNRAS.483.5583V}. We generate 500 lens systems such that the stellar density~($\Sigma_*$) at the global-minima is always greater than~$50~{\rm M}_\odot/{\rm pc^2}$. For our sample, the median value of macro-magnification~($\mu$) is~$\simeq3.8$. The lower limit on the~$\Sigma_*$ makes sure that the source does not lie very far from the diamond caustic in the source plane; otherwise, the lensing magnification and stellar density will drop further, effectively equivalent to no strong lensing. After that, we generate the stellar mass microlenses following the Salpeter initial mass functions~\citep[IMF;][]{1955ApJ...121..161S} in the mass range~$[0.08,1.5]~{\rm M_\odot}$ in a circular region with a radius of~$6~{\rm pc}$. Such a large region is necessary to properly account for microlensing effects at lower frequencies, $f\sim10$~Hz, which corresponds to a time delay of $\sim0.1$~seconds. The microlens mass range is chosen while keeping in mind the fact that a typical ETG will only have stars older than a few Gyr, and within this time, all the stars with masses above~$1.5~{\rm M_\odot}$ will complete their life and form remnants~(like WD, NS, and BH). In principle, when generating the total microlens population, one also needs to generate the remnant population considering an initial-final mass relation as done in earlier work~\citep[e.g.][]{2022MNRAS.517..872M}. However, we do not include remnant population in our current work as such a population will mostly increase the number of microlenses with mass~${\lesssim}0.5~{\rm M_\odot}$ and are not expected to introduce any significant effects in lensed GW signal. The above assumption will not be true if a massive remnant lies near the strongly lensed image at the patch centre. However, such cases can be treated as lensing by a massive remnant in the presence of stellar microlenses, and again our results will be applicable since it is equivalent to lensing by an IMBH in the presence of stellar microlenses.

Results corresponding to global minima microlensing by stellar population are shown in Figure~\ref{fig:Ff_stel}. The solid black curve in the left panel shows the average~(over 500~realisations) of the absolute value of amplification factor~$|F|$ normalised by the corresponding macro-amplification~($\sqrt{\mu}$), i.e.,$|F|/\sqrt{\mu}$. The shaded region covers the~$\pm3\sigma$ range around the average. The blue and yellow curves correspond to two individual realisations. The middle panel represents the relative difference between macro-amplification and amplification factor in the presence of stellar microlenses. The red dashed curve represents the best-fit power law with an index of~0.51 to the average relative difference in frequency range~$f\sim[10^2,10^3]$~Hz. The right panel represents the mismatch for equal mass binaries between unlensed and lensed GW signals as a function of total binary mass.

From the left and middle panels in Figure~\ref{fig:Ff_stel}, we notice that the average relative difference between strong lensing macro-amplification factor~($\sqrt{\mu}$) and amplification factor in the presence of stellar microlenses is~$<1\%$ at~$f=10$~Hz, and~$<2\%$ at~$f=10^2$~Hz. Even considering the~$3\sigma$ range around the average difference at~$f=10$~Hz is~$<2\%$. This implies that, at low frequencies, the effect of stellar microlenses is negligible, and the amplification factor~$F(f)$ can be considered equal to the macro-amplification factor,~$\sqrt{\mu}$. This is further confirmed by the match plot in the right panel where the match between unlensed~(or strongly lensed signal) and microlensed signal, $\mathcal{M}_{\rm Stel}$, is greater than~99.95\% in nearly~95\% of the realisations. Here we use \textsc{PyCBC}~\citep{2016CQGra..33u5004U} with IMRPhenomPv2 approximant to calculate match values as done in~\citet{2022MNRAS.517..872M}. Such large match values imply that in the presence of stellar microlenses only, we can study the global minima assuming negligible microlensing effects due to stellar microlenses. Our results for stellar microlensing agree very well with~\citet{2022MNRAS.517..872M}. In addition, we also note a trend in the match values~(obvious from the background pixel colors) that for binaries with smaller total mass~($M_{\rm Tot}\sim 20~{\rm M_\odot}$), we observe more systems with smaller match~$(<99.95\%)$. This can be understood from the fact that smaller binaries can emit GWs at higher frequecies increasing their chances to get affected by stellar microlenses.

\begin{figure*}
    \centering
    \includegraphics[scale=0.55]{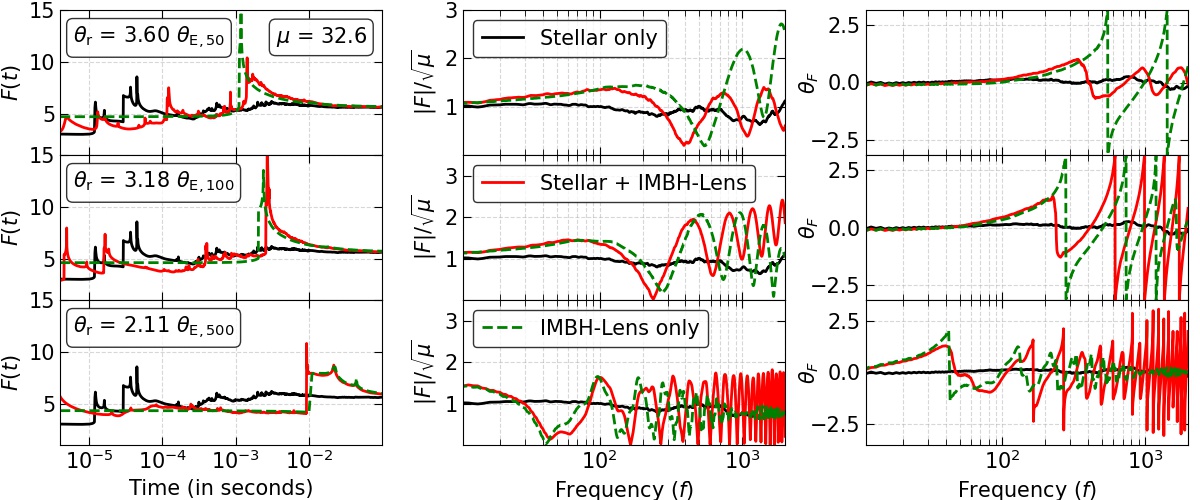}
    \caption{An example~(R8) of microlensing effects in highly magnified global minima similar to Figure~\ref{fig:Ff_stelBH}.}
    \label{fig:Ff_stelBH_Higher_39}
\end{figure*}

\begin{figure}
    \centering
    \includegraphics[scale=0.49]{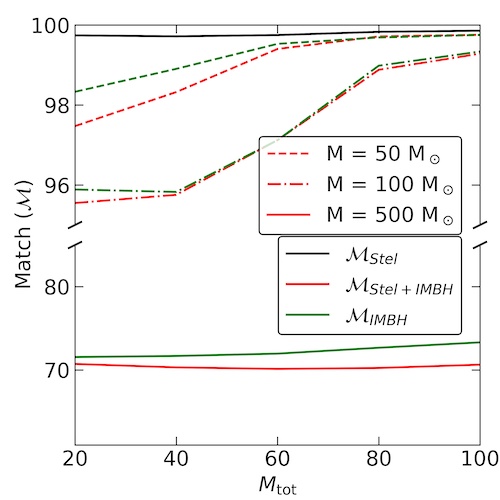}
    \caption{Average effect of stellar microlenses with IMBH on match~($\mathcal{M}$) values for highly magnified global minima GW signals. The black, red, and green curves represent the median match~($\mathcal{M}$) values for stellar microlens only, stellar microlens with IMBH lens, and IMBH only case, respectively, based on the ten realisations. The dashed, dashed-dotted, and solid curves are corresponding to the IMBH mass of~$50~{\rm M_\odot}$, $100~{\rm M_\odot}$, and~$500~{\rm M_\odot}$, respectively.}
    \label{fig:MatchEQM_stelBH_Higher_1}
\end{figure}

\section{Stellar Microlenses $+$ IMBH}
\label{sec:stel_pbh_ml}

As we know, an isolated IMBH leads to the formation of two images~\citep[e.g.,][]{2022JCAP...07..022B}, whereas an IMBH near global minima can form two or four images, depending on its mass and distance from the global minima~\citep[e.g.,][]{2021MNRAS.508.4869M}. If we further add stellar microlenses near global minima, they can affect the position and magnification of images formed due to IMBH. From previous Section, one can naively expect that stellar microlenses cannot affect the match~($\mathcal{M}$) value more than 0.05\% and any additional decrease in~$\mathcal{M}$ is only due to the presence of an IMBH. However, this is not necessarily true as the lens equation governing the image positions is non-linear, and even the stellar microlenses may lead to additional considerable effects in IMBH-led lensing.

To understand the combined effects of IMBH and stellar microlenses on lensed GW signals corresponding to global minima, we randomly place one IMBH in the patch with mass 50~${\rm M_\odot}$, 100~${\rm M_\odot}$, and 500~${\rm M_\odot}$ such that the patch centre~(i.e., position of global minima) lies within the five~Einstein radius of the IMBH. In Figure~\ref{fig:Ff_stelBH}, we show one example of the combined effects of stellar microlenses and IMBH on amplification. The top, middle, and bottom rows, the IMBH mass is~50~${\rm M_\odot}$, 100~${\rm M_\odot}$, and 500~${\rm M_\odot}$, respectively. In each row, the left, middle, and right panels show the Fourier transform~($F(t)$), absolute value~($|F|$), and phase~($\theta_F$) of amplification factor, respectively. In each panel, the black curve is for the stellar microlenses only case, the red curve is for stellar microlenses, along with the IMBH lens, and the green-dashed curve is for the IMBH only case. In~$F(t)$ curves, each spike corresponds to a saddle-point micro-image, and each discontinuity corresponds to a minima micro-image. We refer reader to~\citet{1995ApJ...442...67U, 2021MNRAS.508.4869M} for more details on~$F(t)$ curves.

In the left panels, we notice that all three curves converge to the same value at~$t>0.2$~seconds. The corresponding $F(t)$-value at~$t>0.2$~seconds is equal to the macro-amplification~($\sqrt{\mu}$) value. The prominent spikes in red and green curves at~$t>10^{-3}$~seconds are images corresponding to the IMBH lens. These peaks are responsible for any wave effects at~$f\lesssim10^2$~Hz in~$|F|$ in the middle column. The difference between green and red peak values and positions implies that the presence of stellar microlenses can affect the amplification and position of the IMBH-led images. At~$t<10^{-3}$~seconds, the presence of an IMBH significantly affects the stellar micro-images, which can be seen by the drastic difference between the peak position and heights in the red curve compared to the black curve. Such behaviour implies that at high frequencies~($f\gtrsim10^3$~Hz), the microlensing effects again will have a significant dependence on the IMBH lens in addition to stellar microlenses. In this particular example, shown in Figure~\ref{fig:Ff_stelBH}, we do not see the formation of bright micro-images at~$t<10^{-3}$~seconds. However, from figure~2 in~\citet{2022MNRAS.517..872M}, we can see that stellar microlenses do have the ability to form very bright microimages at~$t<10^{-3}$~seconds. That said, even if stellar microlenses lead to bright images at~$t<10^{-3}$~seconds, it is very unlikely that we will observe any microlensing effects at~$f\lesssim10^2$~Hz as these micro-images will predominantly affect~$f\gtrsim10^3$~Hz which can be seen from~$|F|$ curves in figure~2 of~\citet{2022MNRAS.517..872M}.

Since, at~$t>10^{-2}$~seconds, the formation of microimages is solely determined by the IMBH lens as the stellar microlenses cannot lead to image formation at such large time-delays, the amplification factor at~$f<10^2$~Hz will be completely determined by the IMBH lens. This can be seen in the middle, and right columns of Figure~\ref{fig:Ff_stelBH} as red and green curves overlap with each other at~$t>10^{-2}$~seconds. To quantify the difference between IMBH and IMBH with stellar microlenses, we again calculate the match between lensed and unlensed GW signals. We again consider equal mass component binaries. The corresponding results are shown in Figure~\ref{fig:Ff_stelBH_Match}. In the left and middle columns, we show match~($\mathcal{M}$) as a function of total binary mass for IMBH lens with~($\mathcal{M}_{\rm Stel+IMBH}$) and without stellar microlenses~($\mathcal{M}_{\rm IMBH}$), respectively. Again, in each pixel, the black-coloured values represent the percentage fraction of systems (out of 500) with total binary mass and match shown on the x- and y-axis. Here we only focus on systems with~$\mathcal{M}>95\%$ as smaller match values can only be obtained in cases where IMBH effects in GW signals significantly dominate the stellar microlensing signals. We notice that in nearly~70\% of cases, the match is larger than~99\%. In the~$|\mathcal{M}_{\rm Stel+IMBH}-
\mathcal{M}_{\rm IMBH}|$ histogram plots~(right column), we notice that in nearly 70\%~(90\%) of systems, match only changes by~$<0.1\%~(<0.5\%)$ between IMBH only and IMBH with stellar microlenses cases. Again we do note a trend in the pixels colors showing smaller match value for low mass binaries mostly for~$M_{\rm IMBH} = 50~{\rm M_\odot}$ but it becomes less obvious for other IMBH masses.

Comparing this with~$\mathcal{M}_{\rm Stel}$ values for stellar microlenses only cases shown in the last section, we find that stellar microlenses have the ability to modify the match value by more than~0.1\%. This can be explained by the fact that although stellar microlenses cannot lead to micro-images at~$t>10^{-2}$~seconds, but they can affect the position and amplitude (or magnification) of IMBH-led images leading mismatch value~$>0.1\%$. However, still the overall behaviour of the amplification factor at~$f<10^2$~Hz closely follows the IMBH lensing, as we can see from Figure~\ref{fig:Ff_stelBH} and~\ref{fig:Ff_stelBH_Match}.

\section{Highly magnified global minima}
\label{sec:high_mag}

In a typical case of galaxy lensing, the global minima image is only magnified by moderate macro-magnification factors. For example, in global minima cases simulated in the above section(s), the macro-magnification~($\mu$) was always smaller than~20 with $3.8$ as the median value. However, if the source lies near the cusp point on the major axis of the lens or if the lens ellipticity is small and the source is giving rise to a cross-like image formation (with a large value of tangential shear), even the global minima can be highly magnified. Although such cases are rare but are very interesting as the microlensing effects are very sensitive to~$\mu$ value~\citep[e.g.,][]{2021MNRAS.508.4869M}. The increase in~$\mu$ value also increases the area covered by different time-delay contours. Hence, to correctly cover the contour of $0.1$~seconds time-delay, we needed to increase the patch radius from 6~pc to 10~pc. Keeping in mind the larger patch (implying more expansive simulations), to study such cases, we simulated ten global minima realisations with macro-magnification~($\mu$) in the range~[20, 50]\footnote{Increasing the macro-magnification further means simulating an even larger patch, significantly increasing the computation cost. In such cases, instead of simulating a circle, the better choice is a rectangle~(or an ellipse) scaled according to macro-magnification values. Devising an optimal way to simulate such high magnification cases is subject to our ongoing research, and the results will be discussed in future work.}.

The amplification factor curves for one realisation (R8; with macro-magnification of~32.6) are shown in Figure~\ref{fig:Ff_stelBH_Higher_39} and the rest of the realisations are shown in Appendix~\ref{sec:high_app}. Different curves are the same as Figure~\ref{fig:Ff_stelBH}. The first difference from global minima with typical~$\mu$ values is the formation of many more stellar micro-images~(which can be seen in~$F(t)$ curves) in both black and red curves at~$t<10^{-3}$~seconds. The other important difference is the shift in time delay values of IMBH-caused images at~$t>10^{-3}$ seconds in red curves compared to green curves, which becomes more frequent in high macro-magnification cases. Such a shift mainly affects the~$F(f)$ by changing the positions of crest and troughs in the interference pattern, as we can see in the corresponding~$|F|$ and~$\theta_F$ curves. As mentioned above, the amplitude of these oscillatory features is determined by the peak values in~$F(t)$ curves. At high $\mu$ values, IMBH~($M=500~{\rm M_\odot}$) is able to form bright micro-images at large time~$t>10^{-2}$~seconds which were only appearing rarely in the typical macro-magnification cases discussed in Section~\ref{sec:stel_pbh_ml}. The presence of such images results in a significant decrease in the~$|F|$ at~$f\sim20-30$~Hz as we can see in the bottom row of Figure~\ref{fig:Ff_stelBH_Higher_39}. That said, we still see overall the same behaviour in the~$|F|$ and~$\theta_F$ curves corresponding to IMBH lens with and without stellar microlenses at low frequencies,~$f<10^2$~Hz. In addition, at~$f\sim10$~Hz, $|F|$ corresponding to the IMBH lens with and without stellar microlenses have similar values, implying that at these frequencies, the IMBH lens mainly governs the microlensing effects. That said, once we add an IMBH lens, We can observe differences between~$|F|$ curves for stellar microlenses with IMBH and IMBH only (for example see R6 and R9 in Appendix~\ref{sec:high_app}). This can be explained if we look at the corresponding~$F(t)$ curves where the presence of stellar microlenses can shift and modulate the IMBH-led peaks, which in turn affect the~$|F|$ behaviour at low frequencies. Hence, even though stellar microlenses alone are not enough to give significant effects at~$f<100$~Hz, they can still have non-zero effects in the presence of an IMBH mass lens.

The median match~($\mathcal{M}$) values for these high-$\mu$ cases are shown in Figure~\ref{fig:MatchEQM_stelBH_Higher_1} whereas match values for individual systems are shown in Figure~\ref{fig:MatchEQM_stelBH_Higher_2}. In Figure~\ref{fig:MatchEQM_stelBH_Higher_1}, black, red, and green solid curves show the median match value for stellar microlenses only cases~($\mathcal{M}_{\rm Stel}$), stellar microlenses with IMBH lens~($\mathcal{M}_{\rm Stel+IMBH}$), and IMBH lens only cases~($\mathcal{M}_{\rm IMBH}$), respectively. The dashed, dashed-dotted and solid curves are corresponding to IMBH lens mass is~$50~{\rm M_\odot}$,~$100~{\rm M_\odot}$, and~$500~{\rm M_\odot}$, respectively. As discussed in Section~\ref{sec:stel_ml}, for black curve, the macro-magnification is~$<20$ and the corresponding median value of $\mathcal{M}_{\rm Stel}$ is~$\sim99.7\%$. From the median values of~$\mathcal{M}_{\rm Stel+IMBH}$ and~$\mathcal{M}_{\rm IMBH}$, we again notice that $\mathcal{M}$ values depends on the IMBH lens mass considerably~(similar to~$\mu<20$ cases) as we see very high $\mathcal{M}$~values for~$M_{\rm IMBH}=50~{\rm M_\odot}$ compared to $M_{\rm IMBH}=500~{\rm M_\odot}$. This can be understood (also discussed in last section) from the fact that for a point mass lens, the time delay is proportional to its mass; hence, an IMBH lens with mass $500~{\rm M_\odot}$ can form images at~$t>10^{-2}$~seconds more easily compared to IMBH with mass~$50~{\rm M_\odot}$. This implies that $500~{\rm M_\odot}$ lens is more efficient in introducing microlensing effects at low frequency, leading to a larger mismatch between unlensed and lensed signal compared to a $50~{\rm M_\odot}$ lens. The formation of bright images at large time delays also depends on the background macro-magnification implying that a massive lens with a large $\mu$ value will introduce more microlensing effects at~$f<10^2$~Hz compared to a relatively small mass microlens with a large $\mu$ value. Although not clear from median values, comparison of~$\mathcal{M}_{\rm Stel+IMBH}$ and~$\mathcal{M}_{\rm IMBH}$ shows that even though stellar microlenses alone mostly lead to match values larger than~$99.5\%$ but they are capable of introducing mismatch larger than~$1\%$ in presence of an IMBH lens as can be seen from Figure~\ref{fig:MatchEQM_stelBH_Higher_2}. Once again, it can be explained by the fact that staller microlenses can modify the peak value and position of IMBH-led images, as we can see in Figure~\ref{fig:Ff_stelBH_Higher_39}. For~$M_{\rm IMBH}=50~{\rm M_\odot}~{\rm and}~100~{\rm M_\odot}$, we notice that for small values of total binary mass, the~$\mathcal{M}$ value is smaller and we go towards massive binaries the~$\mathcal{M}$ value increases. However, for~$M_{\rm IMBH}=500~{\rm M_\odot}$, the median~$\mathcal{M}$ value is nearly constant. This can be understood from the fact that smaller binaries can reach to higher frequencies~($\gtrsim500$~Hz) which can get affected even by the less massive IMBHs such as~$50~{\rm M_\odot}$ whereas massive binaries will emit GWs at relatively lower frequencies and will show higher~$\mathcal{M}$ values. But, as we can see from Figure~\ref{fig:Ff_stelBH_Higher_39} (and plots in Appendix~\ref{sec:high_app}), a~$500~{\rm M_\odot}$ IMBH lens can easily introduce wave effects at frequencies~$<100$~Hz effecting the~$\mathcal{M}$ values even for the massive binaries.

\section{Constraining IMBH population in galaxy lenses}
\label{sec:const_bh_pop}
Since the wave effects introduced by stellar microlenses are nearly negligible in global minima, such behaviour can be used to constrain the population of IMBH lenses in the lens galaxy itself. To do so, we need to calculate the probability of an IMBH lying within a certain distance from the macro-image such that the IMBH can introduce detectable wave effects in the signal. As discussed in~\citet{2018PhRvD..98j3022C}, an observed signal-to-noise~(SNR) of~${\sim}30$ is sufficient to detect lensing by an isolated microlens with mass of~${\sim}30~{\rm M_\odot}$ in the LIGO. Similarly, in~\citet{2018PhRvD..98h3005L}, authors find that the LIGO is able to detect isolated IMBHs in mass range~${\sim}[160, 10^3]~{\rm M_\odot}$ with impact parameter in range~${\sim}[0.01, 3]$ and unlensed signal observed SNR of~${\sim}[9, 32]$. Based on the observed GW signal by the LIGO, various studies have tried searching for lensed GW signals and put constraints on the IMBH population~\citep[e.g.,][]{2022ApJ...926L..28B, 2023arXiv230408393T}. However, these studies assumed an isolated IMBH population, whereas, in this work, we focus on the IMBH lensing of strongly lensed GW signals.

We assume that if the macro-image lies with one Einstein radius of the IMBH~(i.e.,~$y\leq1$), we can detect the microlensing effect in the GW signal. This assumption is well justified from observations of amplification factor,~$F(f)$, above where microlensing effects are mainly determined by the IMBH lens even when~$y<5$ and from earlier microlensing studies such as~\citet{2018PhRvD..98h3005L}. Since observing the IMBH lensing of the global minima is a Poisson process, the probability that a given macro-image will be further affected by an IMBH is given by
\begin{equation}
P_l(\tau) = 1-e^{-\tau},
\end{equation}
where~$\tau$ is the IMBH lensing optical depth. For our case, the IMBH lensing optical depth is essentially the fraction of the 
area covered by the Einstein radii of IMBHs, 
\begin{equation}
    \tau = N_{\rm IMBH} \times f(A_{\rm {\theta_E}} (\mu)),
\end{equation}
where~$N_{\rm IMBH}$ represents the number of IMBHs and $f(A_{\rm {\theta_E}} (\mu))$ is the fraction of area covered by the Einstein radius of one IMBH. Here, we can see that the optical depth is a function of macro-magnification and the fraction of dark matter in the form of IMBH~\citep[e.g.,][]{2022ApJ...926L..28B, 2022MNRAS.509.1358U, 2023MNRAS.518..149Z}. Here, we assume a monochromatic mass function for IMBH in the mass range~$[50,10^3]~{\rm M_\odot}$ and~$\mu=3.8$ (which is equal to median magnification in Section~\ref{sec:stel_ml}) as we are only focusing on global minima. Now, we are in a position to ask the following question: How many strongly lensed macro-minima do we need to detect so that we can constrain the fraction of IMBHs to 10\% and 1\% levels with 90\% probability? In such case, the probability that at least one of the macro-minima will be further affected by an IMBH is given by~$1-e^{-N_{\rm event}\tau}$ where~$N_{\rm event}$ is the number of observed GW signals corresponding to macro-minima. From here, a straightforward calculation with the above numbers leads us to the result that if more than~$10\%~(1\%)$ of dark matter is made of IMBHs in the above mass range, then the probability that at least one out of every~${\simeq}15~({\simeq}150)$ of global minima is further lensed by an IMBH is~$\simeq0.9$. This essentially implies that if we do not detect any IMBH microlensing signatures in every~${\simeq}15~({\simeq}150)$ global minima, then we can constrain the fraction dark matter in the form of IMBHs to be less than~$10\%~(1\%)$ with ${\simeq}90\%$ confidence. Here, we can see that the above are sensitive to the macro-magnification. Hence, it is worth checking how the change in median magnification will affect the constraints. To do so, we decrease the median magnification by a factor of two and again perform the same estimates. With this, we find that the number of required lensed GW signals doubles if we want to get similar constraints on the fraction of the IMBH population. To be more specific, assuming a median magnification of~1.9, we need~${\simeq}33~({\simeq}330)$ lensed GW signal to constrain the fraction of IMBH population to~$10\%~(1\%)$ with ${\simeq}90\%$ probability. This explains one of the reasons why the microlensing-only studies looking for isolated such compact objects in the Universe require more lensed GW detections to achieve the same level of constraints on their population since they will have a strong lensing magnification factor of one. However, this may not be true if we only consider IMBHs since their number density may be very small in galaxy-scale lenses.

In the above estimations, we have assumed that we are capable of detecting the IMBH lensing (or the corresponding signatures) in the global minima with IMBH being in the mass range~$[50,10^3]~{\rm M_\odot}$. From~\citet{2018MNRAS.476.2220L}, we can see that for IMBH mass~${>}150~{\rm M_\odot}$, LIGO is able to detect microlensing signature for SNR values~${>}9$. In addition, as the stellar microlensing effects are negligible, for IMBH masses~${>}150~{\rm M_\odot}$, the above assumption is justified. For IMBH masses~${\sim}50~{\rm M_\odot}$, the SNR required to segregate microlensing effects will be~${\sim}30$~\citep{2018PhRvD..98j3022C}. Hence, to probe such IMBH masses, we will require more lensed signals than quoted above. We need to perform a detailed parameter estimation analysis to determine how many global minima will be required to detect these IMBH masses (along with the required SNR values) in galaxy scale lenses. We leave this for future analysis.

\section{Discussion \& Conclusions}
\label{sec:conclusions}

In our current work, we have studied the microlensing effects due to stellar microlenses along with an IMBH on the GW signals corresponding to global-minima images in galaxy scales lenses. We simulate 500 systems corresponding to global minima images with macro-magnification $<20$ and stellar density~$>50~{\rm M_\odot}$. For each system, we simulate four different realisations, one with only stellar mass microlenses in range~$[0.08, 1.5]~{\rm M_\odot}$ and three realisations with stellar mass microlenses along with an IMBH lens with mass values~$50~{\rm M_\odot}$, $100~{\rm M_\odot}$, and $500~{\rm M_\odot}$. We find that, in the absence of an IMBH lens, stellar microlenses lead to negligible microlensing effects in the lensed GW signal. At lower frequencies~$\sim10$~Hz, the relative difference between macro-magnification~($\mu$) and stellar microlensing led amplification factor~($|F|$) is $<2\%$. As we move towards higher frequencies, the differences also increase up to~$\sim10\%$ at frequencies around~$10^2$~Hz. From earlier study~\citet{2022MNRAS.517..872M}, and match study performed in our current work, we conclude that such differences can only give rise to a mismatch~$<0.05\%$ between lensed and unlensed GW signals corresponding to typical global minima. Thanks to the negligible effects introduced by stellar microlenses in~$F(f)$, any differences arising at low frequencies~($f<10^2$~Hz) from macro-amplification~$\sqrt{\mu}$ are solely due to the presence of an IMBH lens near the macro-image. We show this by calculating the difference between match values for isolated IMBH lenses and IMBH lenses along with stellar microlenses. We find that in~${\sim}70\%~({\sim}90\%)$ of realisations out of 500, $|\mathcal{M}_{\rm Stel+IMBH}-\mathcal{M}_{\rm IMBH}|$ is less than~0.1\%~(0.5\%) implying that the additional mismatch is primarily induced due to the presence of an IMBH. To understand the effects of stellar microlenses and IMBHs on highly magnified~(although rare) strongly lensed GW signal, we simulate ten realisations with~$\mu\in[20, 50]$ range. We again observe that~$|F|$ and~$\theta_F$ curves corresponding to IMBH only case and IMBH with stellar microlenses case closely follow each other at~$f<10^2$~Hz implying that microlensing effects at these frequencies are mainly governed by the IMBH lens. This analysis implies that a typical lensed GW signal corresponding to global minima will behave as it is only strongly lensed and can be used to gauge the amplitude of microlensing effects in the other counterparts. A similar conclusion was also reached in~\cite {2022MNRAS.517..872M}, and here we have shown the validity of these results for even higher strong lensing magnification values.

We also notice that the presence of an IMBH with mass~$500~{\rm M_\odot}$ near a highly magnified image can easily lead an image at time delays~$>10^{-2}$~seconds leading to a significant drop in corresponding~$|F|$ curve at~$f\sim30$~Hz. Since stellar microlensing effects are negligible in global minima, such effects can be used to probe the (wandering) IMBH population inside the galaxy scale lenses. We find that assuming a monochromatic mass function for IMBH and a typical macro-magnification of~3.8 for global minima, non-detection of microlensing effects in~${\simeq}15~({\simeq}150)$ global minima can rule out~$10\%~(1\%)$ fraction of dark matter is made of IMBHs with mass values~$>150~{\rm M_\odot}$ inside the galaxy scale lenses whereas for IMBH masses~$>50~{\rm M_\odot}$ the required number of global minima will be larger since the SNR needed to detect lensing effect due to less massive IMBHs will be higher.

Here, it is important to point out that, in our current work, we have specifically focussed on IMBHs~(in mass range~$[50, 10^3]~{\rm M_\odot}$) lensing of global minima as it cannot be de-magnified by strong lensing, and the stellar density is minimum. However, similar effects can be expected for any other lenses (for example, dark subhalos) given that the corresponding effective lensing mass (mass enclosed inside Einstein radius) is similar to IMBHs. That said, differences can be expected in the resulting wave effects if the radial profile of these lenses differs from the point mass. In addition, strong lensing forms multiple copies of the same GW signal and these other counterparts will arrive at us with a certain time delay. As discussed in~\citet{2022MNRAS.517..872M}, for~$|\mu|<10$, these counterparts are also not expected to be affected significantly by stellar microlenses and might also become a probe for IMBH lenses, which lie relatively close to the lens centre where the density of these IMBHs is expected to be higher. That said, 50\% of observed strongly lensed GW signals are expected to have~$|\mu|>10$. Hence, further detailed study is required to check whether these other counterparts will be significantly affected by the stellar microlenses or not. If proven (that stellar microlenses do not introduce significant effect in any of the lensed GW counterparts), lensed GW signals in the LIGO band will become excellent probes for IMBHs (or subhalos) inside the galaxy (and cluster) scale lenses.

\section{Acknowledgements}
AKM would like to thank Anuj Mishra, Anupreeta More, and Jasjeet Singh Bagla for their useful comments. AKM thanks the anonymous reviewers for their useful comments. AKM acknowledges support by grant 2020750 from the United States-Israel Bi-national Science Foundation (BSF) and grant 2109066 from the United States National Science Foundation~(NSF) and the Ministry of Science $\&$ Technology, Israel.
\\
\\
\textit{Software:}
\textsc{python}~(\url{https://www.python.org/}),
\textsc{astropy}~\citep{2018AJ....156..123A},
\textsc{cython}~\citep{2011CSE....13b..31B},
\textsc{matplotlib}~\citep{2007CSE.....9...90H},
\textsc{numpy}~\citep{2020Natur.585..357H},
\textsc{scipy}~\citep{2020SciPy-NMeth}.

\section{Data Availability}
The microlensing simulation data can be made available upon reasonable request to the corresponding author.

\bibliographystyle{mnras}
\bibliography{reference}

\appendix
\section{Highly magnified global minima: Realisations}
\label{sec:high_app}

\begin{figure*}
    \centering
    \includegraphics[scale=0.5]{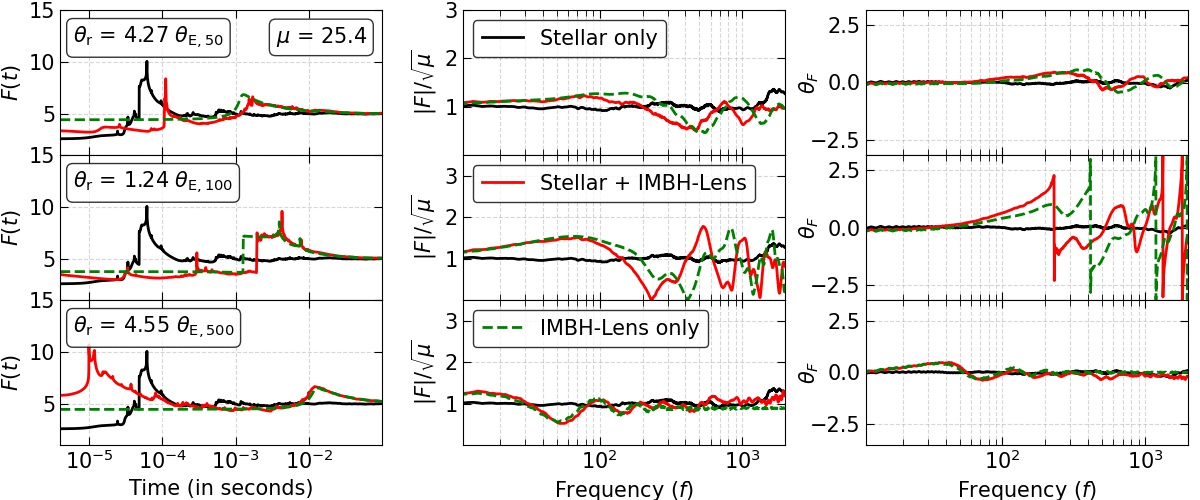}
    \caption{R1: microlensing effects in highly magnified global minima similar to Figure~\ref{fig:Ff_stelBH}.}
    \label{fig:Ff_stelBH_Higher_02}
\end{figure*}

\begin{figure*}
    \centering
    \includegraphics[scale=0.5]{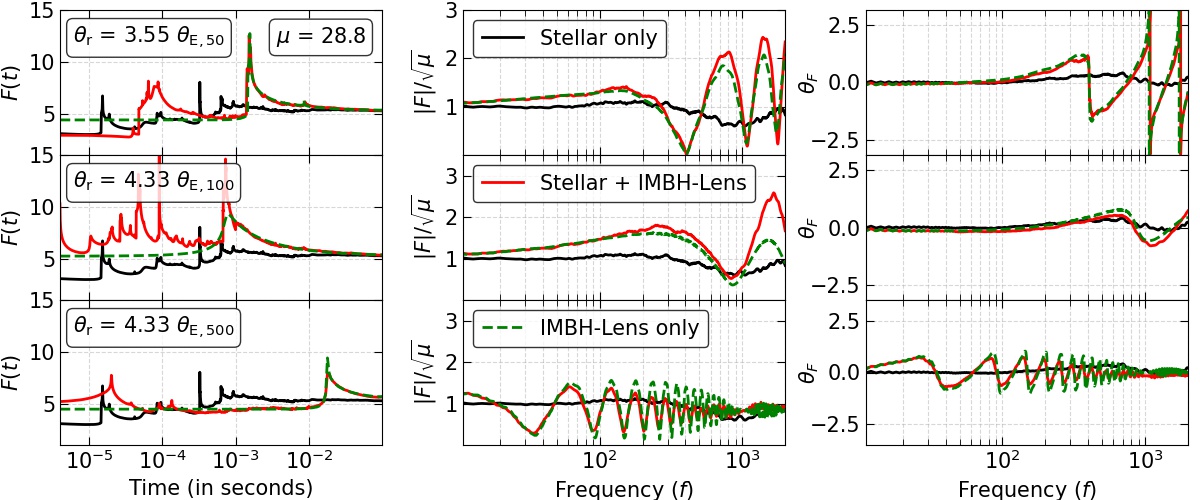}
    \caption{R2: microlensing effects in highly magnified global minima similar to Figure~\ref{fig:Ff_stelBH}.}
    \label{fig:Ff_stelBH_Higher_04}
\end{figure*}

\begin{figure*}
    \centering
    \includegraphics[scale=0.5]{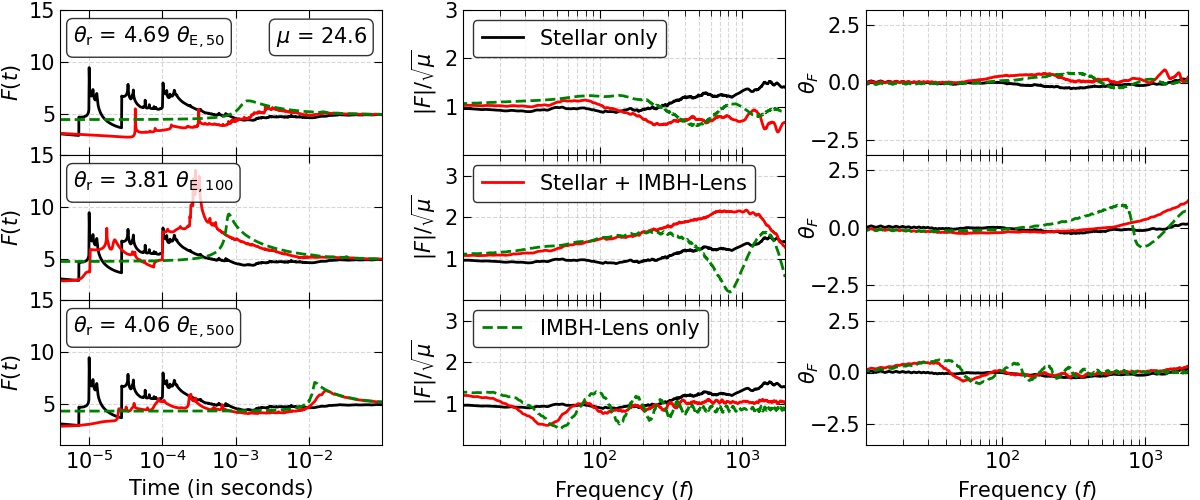}
    \caption{R3: microlensing effects in highly magnified global minima similar to Figure~\ref{fig:Ff_stelBH}.}
    \label{fig:Ff_stelBH_Higher_05}
\end{figure*}

\begin{figure*}
    \centering
    \includegraphics[scale=0.5]{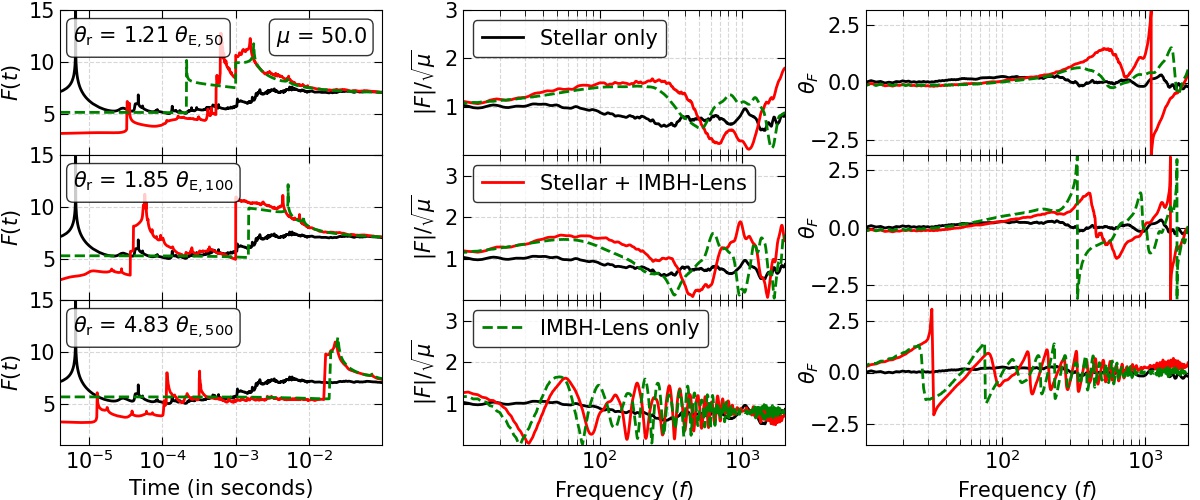}
    \caption{R4: microlensing effects in highly magnified global minima similar to Figure~\ref{fig:Ff_stelBH}.}
    \label{fig:Ff_stelBH_Higher_18}
\end{figure*}

\begin{figure*}
    \centering
    \includegraphics[scale=0.5]{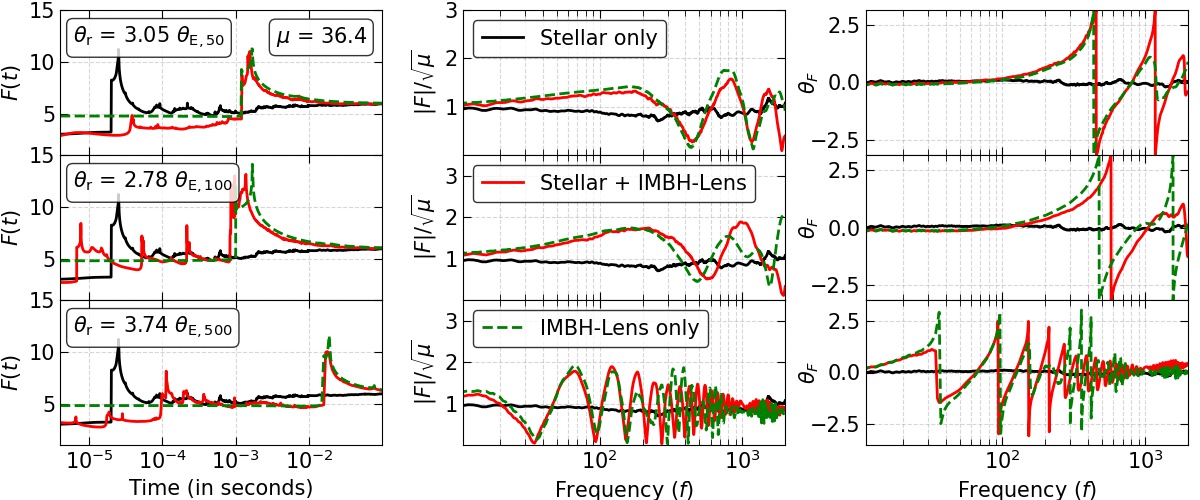}
    \caption{R5: microlensing effects in highly magnified global minima similar to Figure~\ref{fig:Ff_stelBH}.}
    \label{fig:Ff_stelBH_Higher_27}
\end{figure*}

\begin{figure*}
    \centering
    \includegraphics[scale=0.5]{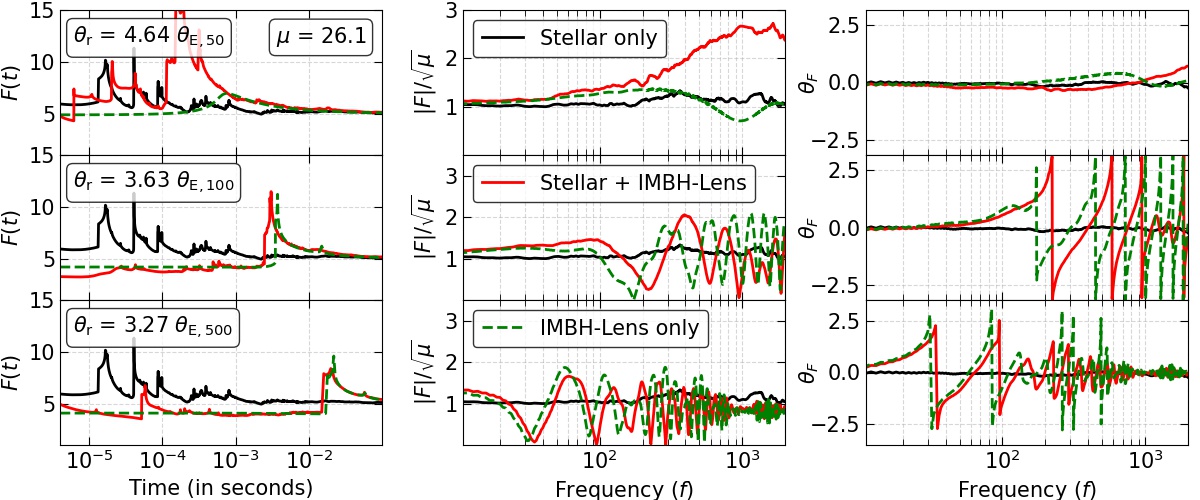}
    \caption{R6: microlensing effects in highly magnified global minima similar to Figure~\ref{fig:Ff_stelBH}.}
    \label{fig:Ff_stelBH_Higher_28}
\end{figure*}

\begin{figure*}
    \centering
    \includegraphics[scale=0.5]{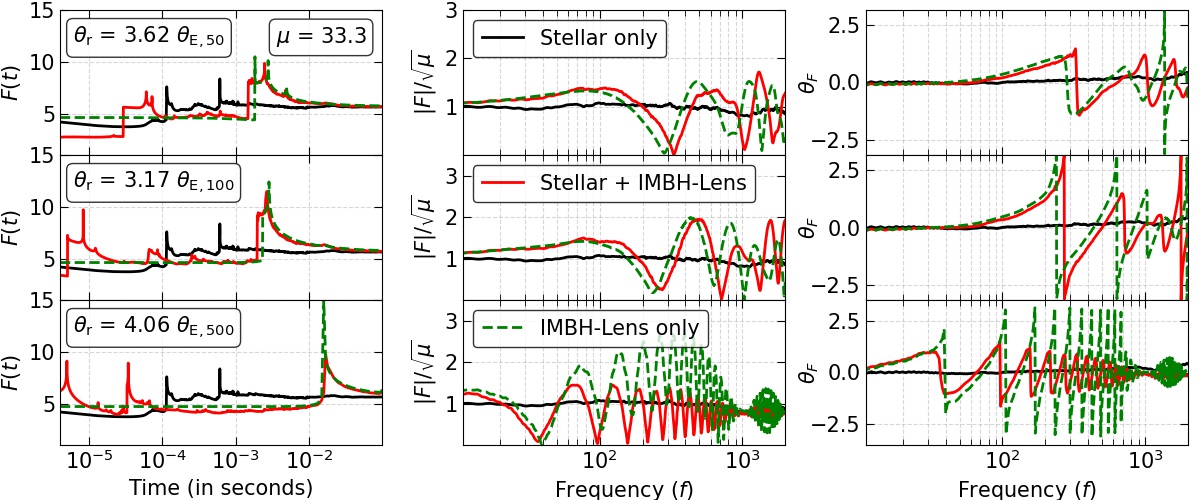}
    \caption{R7: microlensing effects in highly magnified global minima similar to Figure~\ref{fig:Ff_stelBH}.}
    \label{fig:Ff_stelBH_Higher_33}
\end{figure*}

\begin{figure*}
    \centering
    \includegraphics[scale=0.5]{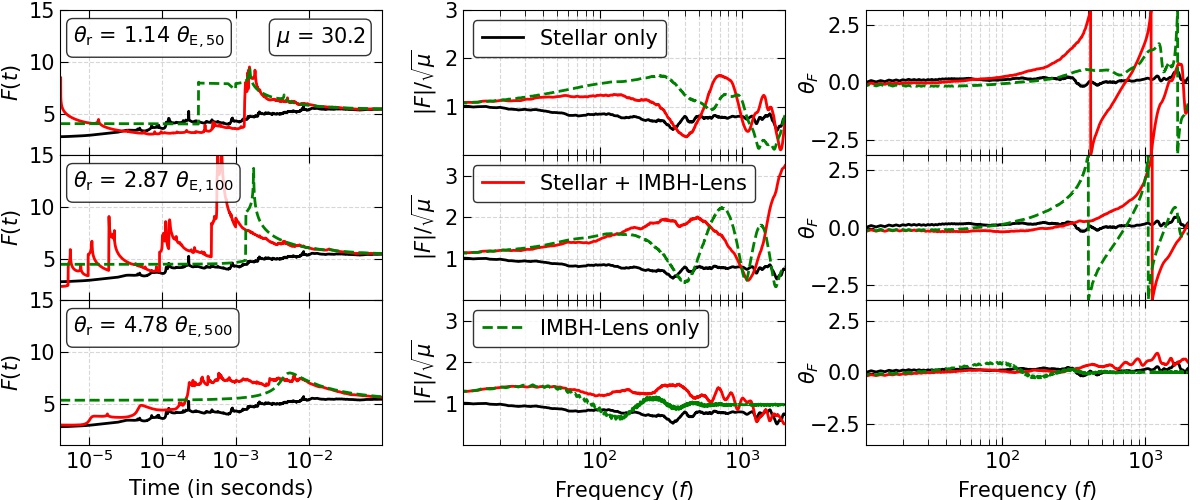}
    \caption{R9: microlensing effects in highly magnified global minima similar to Figure~\ref{fig:Ff_stelBH}.}
    \label{fig:Ff_stelBH_Higher_48}
\end{figure*}

\begin{figure*}
    \centering
    \includegraphics[scale=0.5]{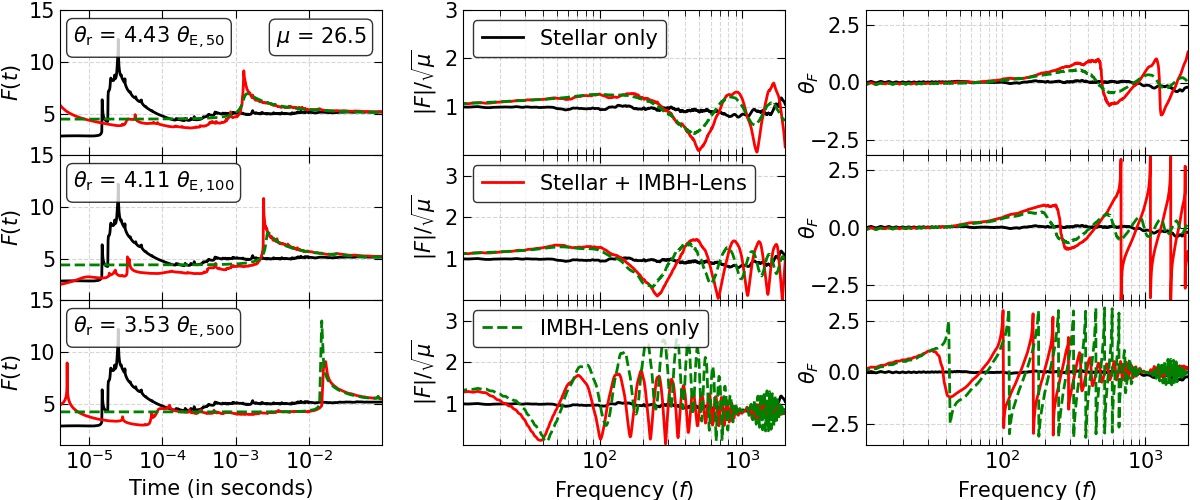}
    \caption{R10: microlensing effects in highly magnified global minima similar to Figure~\ref{fig:Ff_stelBH}.}
    \label{fig:Ff_stelBH_Higher_50}
\end{figure*}

\begin{figure*}
    \centering
    \includegraphics[height=10cm,width=17cm]{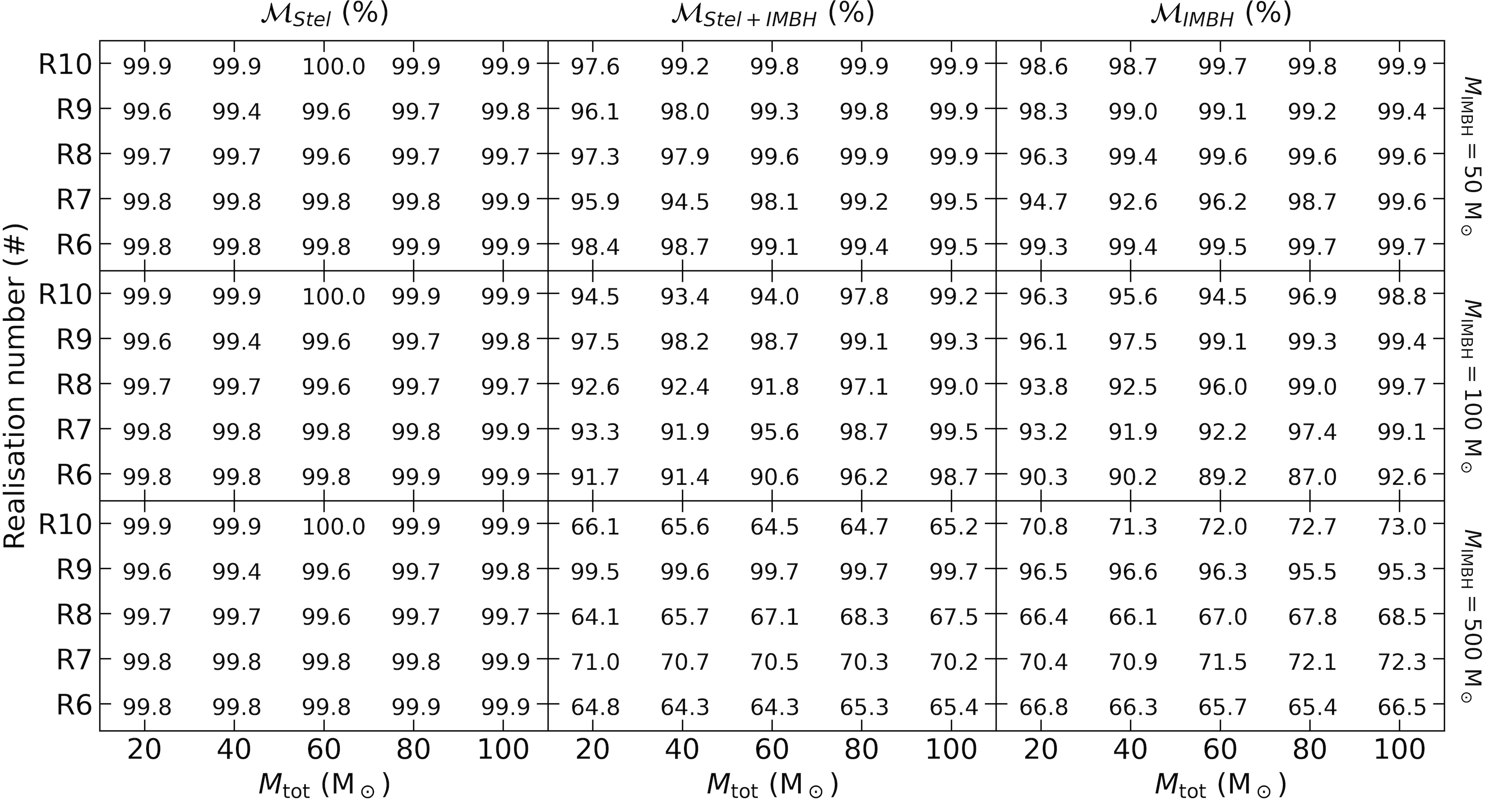}
    \includegraphics[height=10cm,width=17cm]{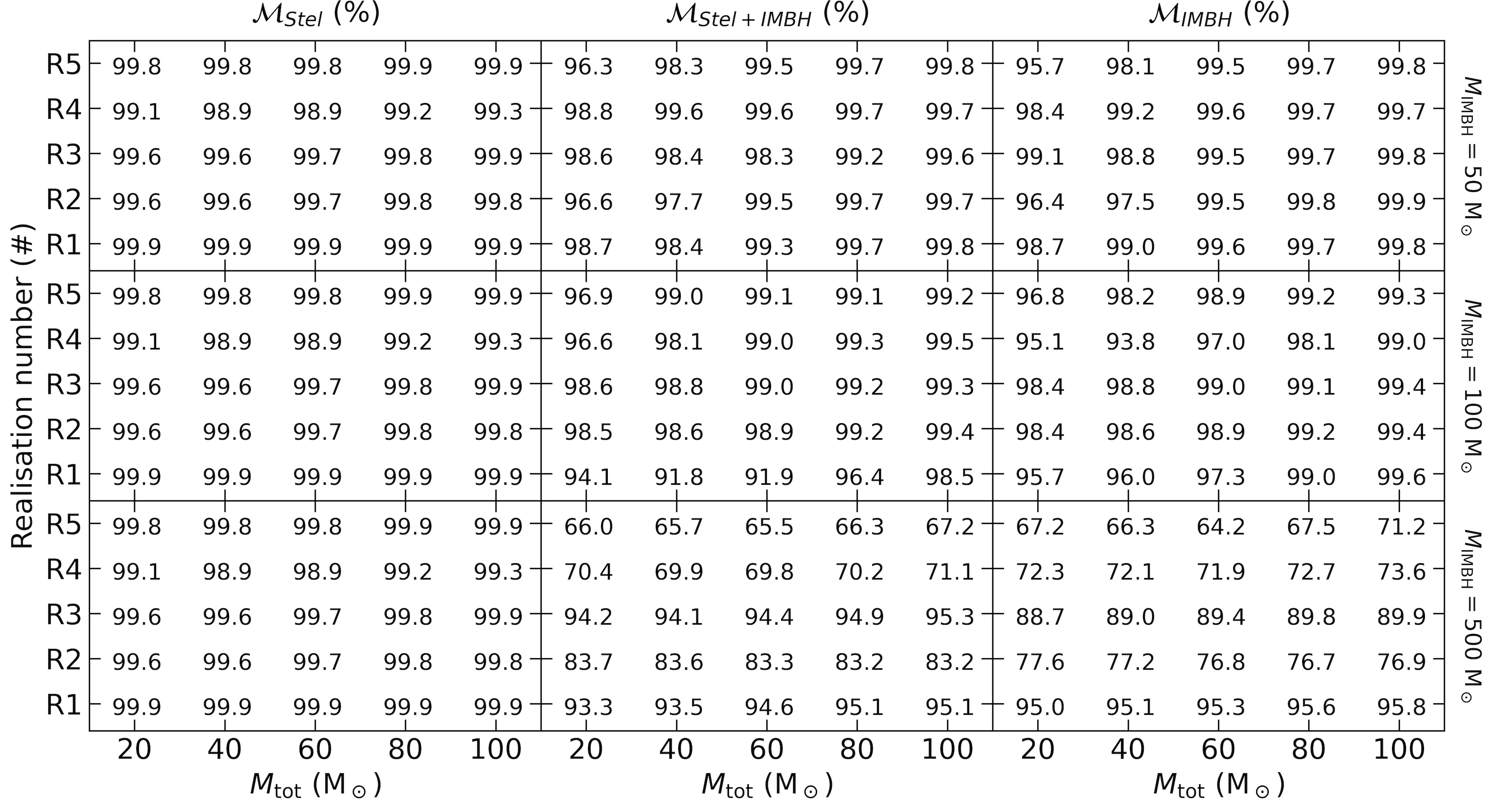}
    \caption{Effect of stellar microlenses with IMBH on match~($\mathcal{M}$) values for highly magnified global minima GW signals. In the bottom and top plots, we show the values for the first to fifth and sixth to tenth realisations. In each panel, the x-axis is the total binary mass, and the y-axis is the realisation number. The left, middle, and right columns show the~$\mathcal{M}$ values for stellar microlenses only case, stellar microlenses with IMBH case, and IMBH only case, respectively. In the top, middle, and bottom rows, the IMBH mass is~$50~{\rm M_\odot}$, $100~{\rm M_\odot}$, and~$500~{\rm M_\odot}$, respectively. A match value of~100\% is a result of rounding the values up to one decimal place.}
    \label{fig:MatchEQM_stelBH_Higher_2}
\end{figure*}

\bsp	
\label{lastpage}
\end{document}